\documentclass{article}

\usepackage[utf8]{inputenc}
\usepackage{balance}
\usepackage{cleveref}
\usepackage{booktabs}
\usepackage{subcaption}
\usepackage{url}
\usepackage{xspace}
\usepackage{amssymb}
\usepackage{float}

\usepackage{lineno}
\usepackage{pythonhighlight}
\usepackage{verbatimbox}

\usepackage{caption}
\usepackage{subcaption}

\newcommand{\ie}{i.e.,\xspace}
\newcommand{\eg}{e.g.,\xspace}

\newcommand{\etal}{et al.\xspace}

\usepackage{todonotes}

\newboolean{showcomments}
\setboolean{showcomments}{true} 
\ifthenelse{\boolean{showcomments}}
{\newcommand{\nb}[2]{
  \fcolorbox{black}{yellow}{\bfseries\sffamily\scriptsize#1}
  {\sf\small$\blacktriangleright$\textit{#2}$\blacktriangleleft$}
 }
 
}
{\newcommand{\nb}[2]{}
 
}

\begin{document}
  


\begin{center}
\Large\textbf{RepoMiner: a Language-agnostic Python Framework to Mine Software Repositories for Defect Prediction}

\normalsize Stefano Dalla Palma, Dario Di Nucci, Damian A. Tamburri

\textit{Jheronimus Academy of Data Science, The Netherlands}
\end{center}



\begin{abstract}

Data originating from open-source software projects provide valuable information to enhance software quality.
In the scope of Software Defect Prediction, one of the most challenging parts is extracting valid data about failure-prone software components from these repositories, which can help develop more robust software.
In particular, collecting data, calculating metrics, and synthesizing results from these repositories is a tedious and error-prone task, which often requires understanding the programming languages involved in the mined repositories, eventually leading to a proliferation of language-specific data-mining software.
This paper presents \textsc{RepoMiner}, a language-agnostic tool developed to support software engineering researchers in creating datasets to support any study on defect prediction.
\textsc{RepoMiner} automatically collects failure data from software components, labels them as \textit{failure-prone} or \textit{neutral}, and calculates metrics to be used as ground truth for defect prediction models. 
We present its implementation and provide examples of its application.

\end{abstract}



\section*{Code Metadata}

\begin{table}[H]
\resizebox{\textwidth}{!}{
\begin{tabular}{ll}
\hline
Current code version & 1.0 \\
Permanent link to repository used for this code version & \url{https://github.com/radon-h2020/radon-repository-miner} \\
Legal Code License   & Apache License, 2.0 (Apache-2.0) \\
Code versioning system used & git \\
Software code languages, tools, and services used & python $>=$ 3.6 \\
Compilation requirements, operating environments & Linux \\
Developer documentation & \url{https://radon-h2020.github.io/radon-repository-miner} \\
\hline
\end{tabular}
}
\end{table}

\section{Motivation and Significance}
Mining software repositories (MSR) is common for researchers to collect and empirically investigate the rich data available in software repositories to uncover interesting and actionable insights about software systems and projects.\footnote{\url{http://www.msrconf.org/}}
These insights are particularly useful in software engineering research to identify best practices for improving software quality, resource allocation, bug prediction, and many more~\cite{chaturvedi2013tools}.

Among these, predicting classes prone to defects using prediction models based on machine learning is a mature research area that strongly relies on MSR techniques and tools.
Building defect prediction models typically consist of mining software archives such as version control and issue tracking systems to generate instances from software components, such as classes and methods. 
Instances are characterized using metrics. For example, \textit{process metrics} measures aspects of the development process, such as the number of files committed together. Besides, \textit{source code metrics} relate to the structural characteristics of source files, as the number of code lines.
Afterwards, software components are labelled as ``failure-prone'' or ``neutral'' and used as ground truth by machine learning classifiers to learn the features (\ie the metrics) that discriminate and predict defects in a specific component.
The limitation of such a method is the labelling process which is challenging and time-consuming.

To tackle this problem, we implemented \textsc{RepoMiner}, a language-agnostic framework that combines failure data acquisition and metrics extraction to ease researchers mining software repositories and create datasets for studies on defect prediction.
\textsc{RepoMiner} does not re-invent the wheel by implementing yet another mining framework. 
Instead, it builds on top of the well-known Python framework \textit{PyDriller}~\cite{Spadini2018} to analyze a repository's history and provides APIs to:

\begin{itemize}
    \item Identify defect-fixing commits, that is, commits that take action to remove a defect.
    \item Identify failure-prone software components across the project's history and label them ``failure-prone'' or ``neutral''.
    \item Extract process and language-specific source code metrics from those components and generate a dataset of observations to train defect prediction models.
\end{itemize}

Researchers can manipulate the extracted data programmatically, quickly export the results to JSON and CSV files, and further extend the framework to support new languages.

It is worth noting that most researchers use already developed tools for data extraction from software engineering repositories, pattern finding, learning, and prediction. Nevertheless, often they implement custom scripts for their mining tasks instead of using the available tools.
Chaturvedi~\etal~\cite{MSRtools2013} have surveyed MSR-related papers to understand the usage of different tools and studied their purpose.
Among them, we have found no tools that support automatic labelling of and metrics extraction from software components for defect prediction regardless of the language. 
Kuhn~\cite{Kuhn2009} presents a lexical approach to retrieve labels from source code automatically. However, it is only a prototype implementation and aims to obtain labels describing components to compare rather than identify that deemed failure-prone.
The most similar to \textsc{RepoMiner} is \textsc{Alitheia Core}, an extensible platform for performing large-scale software engineering and repository mining studies~\cite{Gousios2009}. 
Particularly interesting is its capability to be extended by plug-ins that calculate process and product metrics, which inherit from an abstract implementation of a plug-in interface and only have to provide implementations of 2 methods.
Nevertheless, its scope is much broader than \textsc{RepoMiner}'s, it does not directly tackle the mining of failure-prone components, and it is archived.

In this paper, we present the labelling and metrics extraction implementation provided by \textsc{RepoMiner} and give examples of its application. 
The remainder of this paper is structured as follows.
\Cref{sec:tool} describes the \textsc{RepoMiner}'s architecture.
\Cref{sec:template} illustrates how to use it.
In \Cref{sec:case_study}, the tool is instantiated in the scope of Infrastructure-as-Code defect prediction to show possible extensions to new languages.
\Cref{sec:limitation_extensions} provide a detailed outlook on the current limitation and how the tool can provide new functionalities.
Finally, \Cref{sec:conclusions} concludes the paper and outlines future works.

\section{RepoMiner: Characteristics and Outline}\label{sec:tool}

\textsc{RepoMiner} is a language-agnostic mining framework that helps researchers identify failure-prone components from software repositories and support the creation of datasets for defect prediction.
The tool is open-source and available on GitHub\footnote{\url{https://github.com/radon-h2020/radon-repository-miner}} and the Python Package Index\footnote{\url{https://pypi.org/project/repository-miner//}}.
This section describes the design of \textsc{RepoMiner} and its main APIs.

\begin{figure}[ht]
    \centering
    \includegraphics[width=.7\linewidth]{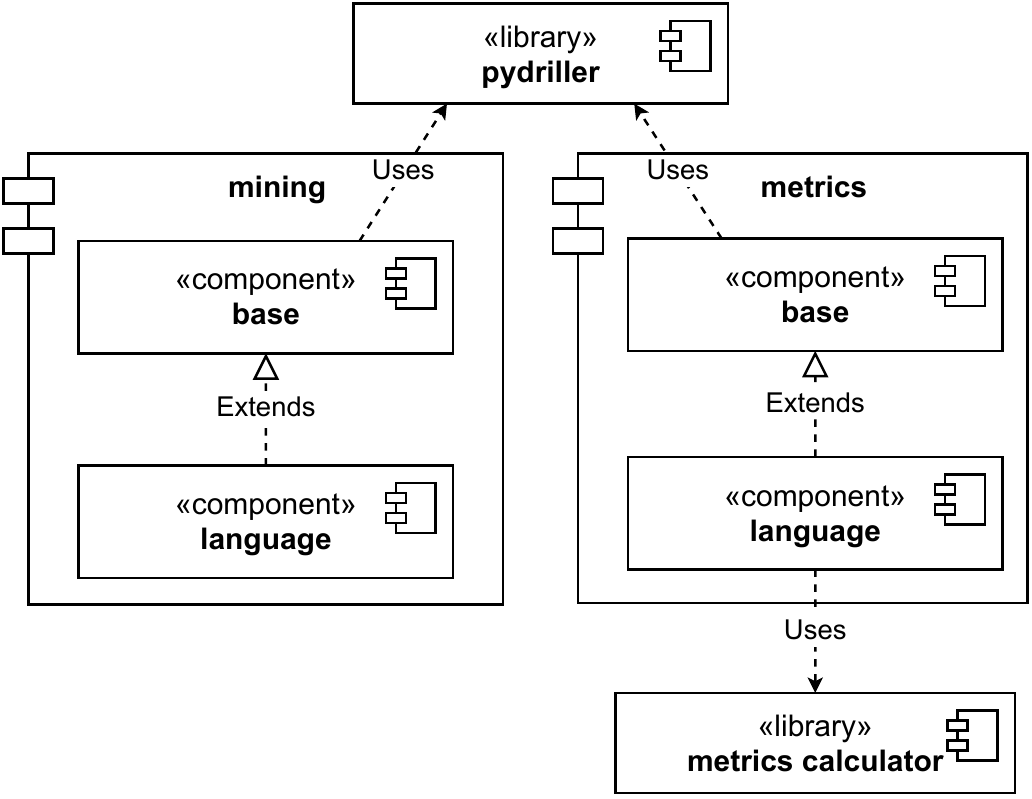}
    \caption{UML component diagram of \textsc{RepoMiner}.}
    \label{fig:agnostic_component_diagram}
\end{figure}

\Cref{fig:agnostic_component_diagram} shows an overview of \textsc{RepoMiner}'s architecture consisting of two modules, namely \texttt{mining} and \texttt{metrics}.
The former builds on top of \textit{PyDriller}~\cite{Spadini2018} to analyze the project history, identify \textit{defect-fixing} and \textit{bug-introducing} commits, and \textit{failure-prone} files needed for the analysis.
The latter relies on \textit{PyDriller} and additional external libraries to extract process and source code metrics from these files.

As its name suggests, a \texttt{base} component provides the base functionalities for that module. 
Support a new language requires creating a \texttt{language} component that extends the functionalities in the \texttt{base} component to fit the language.
Base components implement the classes \texttt{mining.BaseMiner} and \texttt{mining.FixingCommitClassifier} for repository mining and defect-fixing commits categorization, and the class \texttt{metrics.BaseMetricsExtractor} to extract metrics from the mined scripts.
The following sections describe these components in detail.

\begin{figure*}[tb]
    \centering
    \includegraphics[width=0.95\linewidth]{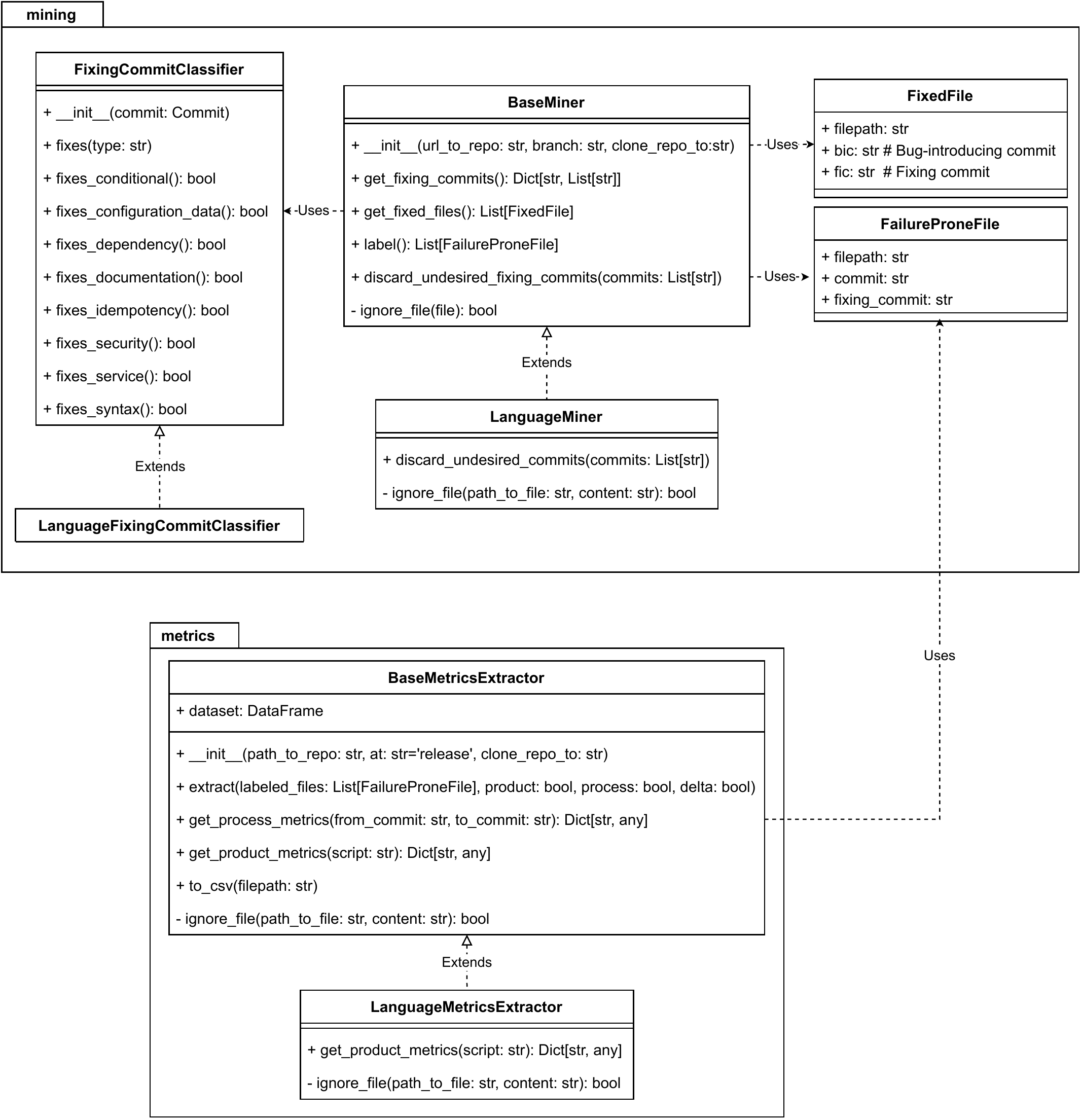}
    \caption{UML class diagram of the \texttt{mining} and \texttt{metrics} modules.}
    \label{fig:class_diagram}
\end{figure*}

\subsection{BaseMiner}\label{subsec:base_miner}

\texttt{BaseMiner} is an abstract class that must be extended to mine specific language-based repositories (\ie \texttt{LanguageMiner} in \Cref{fig:class_diagram}). 
Mandatory inputs is the \textit{URL} to a remote git repository (\eg a GitHub or GitLab repository).
Optional inputs are the \textit{branch} to analyze (\eg \textit{main} or \textit{master}) and the \textit{path to clone} the repository on the disk. 
Then, the mining consists of the three methods below, to be executed sequentially:  \texttt{get\_fixing\_commits()}, \texttt{get\_fixed\_files()}, and \texttt{label()}.

\begin{figure}
    \centering
    \includegraphics[width=0.75\linewidth]{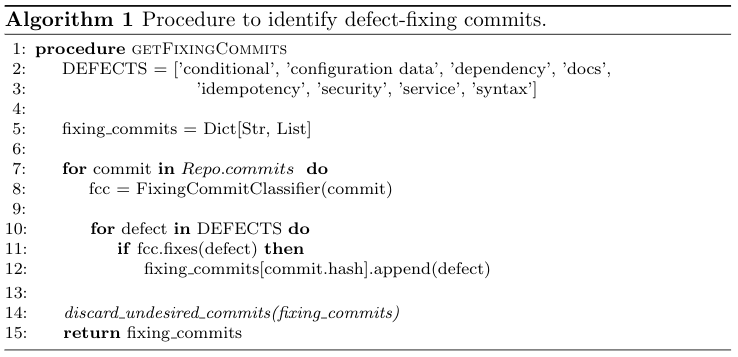}
    \caption{Procedure to identify defect-fixing commits.}
    \label{alg:getFixingCommits}
\end{figure}

\paragraph{Identify defect-fixing commits}
Function \texttt{get\_fixing\_commits()} implements \Cref{alg:getFixingCommits} to identify \textit{defect-fixing} commits (lines 7-12). 
Defects are categorized based on the qualitative analysis performed by Rahman~\etal~\cite{Rahman2020taxonomy} on defect-related commits collected from open-source software repositories of the Openstack organization: ``conditionals'', ``configuration data'', ``dependencies'', ``documentation'', ``idempotency'', ``security'', ``service'' and ``syntax''. 
Given a commit and a defect category, the function identifies whether the commit message or modifications to one of its files indicate a fix, based on the rules defined in~\cite{Rahman2020taxonomy}. 
If so, the commit hash is added to the dictionary of defect-fixing commits with the respective categories (lines 11-12). 
Finally, it keeps only the commits that modify at least one file for the language at hand (line 14) and returns the dictionary of defect-fixing commits.

	

\begin{figure}
    \centering
    \includegraphics[width=0.75\linewidth]{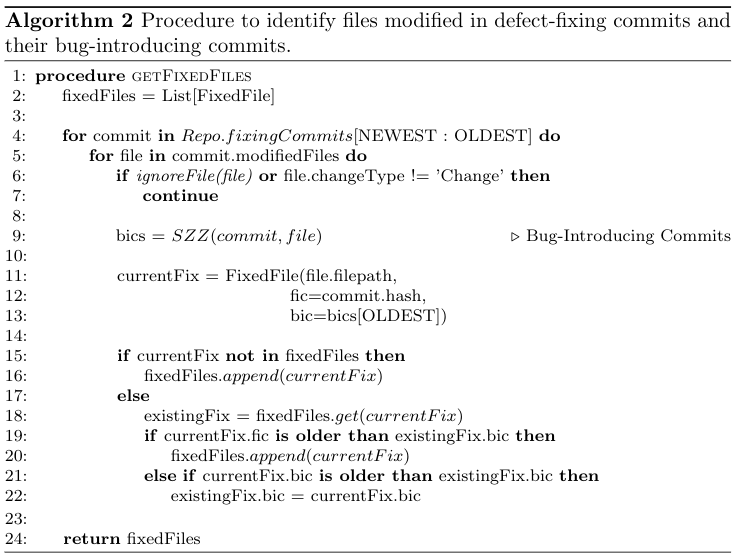}
    \caption{Procedure to identify files modified in defect-fixing commits and their bug-introducing commits.}
    \label{alg:getFixedFiles}
\end{figure}

\begin{figure}[h]
  \centering
  \begin{subfigure}{0.95\linewidth}
    \centering
    \includegraphics[width=0.60\linewidth]{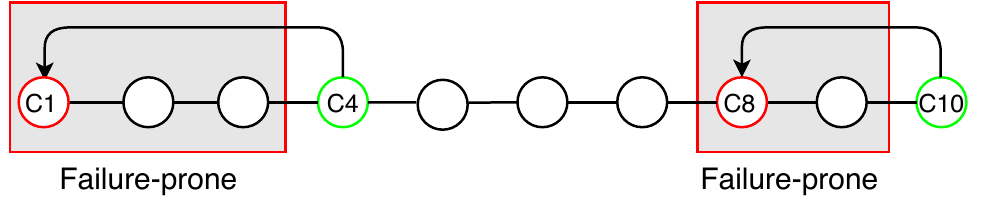}
    \caption{}
  \end{subfigure}
  \begin{subfigure}{0.95\linewidth}
    \centering
    \includegraphics[width=0.60\linewidth]{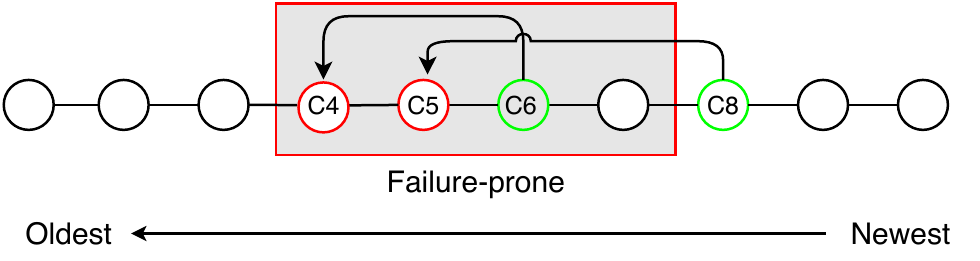}
    \caption{}
  \end{subfigure}
  \caption{Two scenarios of the labeling process where the green and red circles are defect-fixing and bug-introducing commits, respectively.}
  \label{fig:labelling}
\end{figure}

\paragraph{Identify fixed files}
Function \texttt{get\_fixed\_files()} implements \Cref{alg:getFixedFiles} to identify relevant files that are modified in the defect-fixing commits and their \textit{bug-introducing} commit. 
To this end, commits are analyzed backward from the most recent defect-fixing commit to the oldest (line 4).
Given a file, the function ensures that it is written in the language selected for the analysis and it has been modified, that is, it is not a new, deleted, or renamed file (line 6). 
Then, the \textit{SZZ} algorithm\footnote{As implemented in \textit{PyDriller} at release $\geq$ 2.0.}~\cite{kim2006automatic} is used to identify the 
\textbf{oldest} commit that introduced the defect in that file (line 9), known as the \textit{bug-introducing} commit (\textit{bic)}.
The \textit{bic}, along with the defect-fixing commit hash (\textit{fic}) and the \textit{filename}, is used to create a new \texttt{FixedFile} object (line 11).
The file is added directly to the list of fixed files (line 15-16) when it is encountered for the first time through the function.
Successively, the following steps apply:

\begin{itemize}
    \item \textbf{Lines 19-20.} If the current defect-fixing commit (\eg C4 in \Cref{fig:labelling}a) is older than the file's previous \textit{bic} (\eg C8 in \Cref{fig:labelling}a), then a brand new object (\ie \textit{FixedFile(file=A, fic=C4, bic=C1)}) is appended to the list of fixed files. 
    
    \item \textbf{Lines 21-22.} If the file's previous \textit{bic} (\eg C5 in \Cref{fig:labelling}b) is between its current \textit{bic} (\eg C4 in \Cref{fig:labelling}b) and \textit{fic} (\eg C6 in \Cref{fig:labelling}b), the existing \textit{bic} is updated with the current one (\ie from C5 to C4).
\end{itemize}

For the sake of clearness, we omitted the lines handling file-renaming in \Cref{alg:getFixedFiles}.
The function returns the list of \texttt{FixedFile} objects that can be used by the following function to label the files as \textit{failure-prone} at a given point in the repository history. 

\paragraph{Labelling}
Given a list of fixed files, the function \texttt{label()} tags all the snapshots of a \texttt{FixedFile} between its \textit{bic} (inclusive) and its \textit{fic} as \textit{failure-prone}.
Then, for each \texttt{FixedFile} at a given point in the repository history, it yields a \texttt{FailureProneFile} object consisting of the \textit{filepath}, the \textit{commit hash} at that time, and the hash of its \textit{defect-fixing commit}.

It is worth noting that the \texttt{BaseMiner} class in \Cref{fig:class_diagram} has two additional methods, namely \texttt{discard\_undesired\_fixing\_commits()} and \texttt{ignore\_file()}, which are called in \Cref{alg:getFixingCommits} and \Cref{alg:getFixedFiles}, respectively.
These methods \textit{must} be implemented by the sub-classes to filter commits and files for a given language.
The former discards the fixing commits that do not modify files of the language considered.
The latter returns a boolean value indicating whether to ignore a given file based on its path, extension, and/or content. 
For example, if the user aims at analyzing Python-based repositories, the method will ensure that all non \textit{.py} files are ignored.

\subsection{FixingCommitClassifier}
\texttt{FixingCommitClassifier} is an abstract class that categorizes fixing-commits based on the defects categories mentioned in \Cref{subsec:base_miner}. 
It provides base implementations for each category, although some have to be overridden by the classes extending it because of the language.
For example, the methods \texttt{fixes\_configuration\_data()} and \texttt{fixes\_idempotency()} in \Cref{fig:class_diagram} relate to defect categories specific to language for configuration management and infrastructure provisioning and do not apply to application code languages.
By contrast, methods such as \texttt{fixes\_dependency()} are common to different languages, although the implementation can slightly change depending on the syntax.
Therefore, concrete \texttt{LanguageFixingCommitClassifier} classes should be implemented when new languages are added, and related methods overridden.
Finally, an instance of \texttt{LanguageFixingCommitClassifier} must be initialized in the constructor of the respective \texttt{LanguageMiner} to be used in \Cref{alg:getFixingCommits}, as shown in \Cref{fig:python_miner}.

\subsection{BaseMetricsExtractor} 
\texttt{BaseMetricsExtractor} is an abstract class that extracts process and source code metrics from the collected files, and creates a dataset of \textit{failure-prone} and \textit{neutral} observations ready for defect prediction.
It is extended by the classes accountable for extracting source code metrics related to a specific language (i.e., \texttt{LanguageMetricsExtractor} in \Cref{fig:class_diagram}). 

Similarly to \texttt{BaseMiner}, the mandatory input is the \textit{path} to the local or remote git repository, while optional inputs include the repository's \textit{branch} and the \textit{path to clone} the repository in case the mandatory input is a remote URL. 
The entry point is the method \texttt{extract()}. 
It requires a list of \texttt{FailureProneFile} objects and the type of metrics to extract, namely \textit{process}, \textit{product} or \textit{delta}, or group thereof.
Then, methods \texttt{get\_process\_metrics()} and \texttt{get\_product\_metrics()} are used to extract process and source code metrics, respectively.
Please note that, relying on the development activity rather than the source code, process metrics are language-agnostic and do not need any further extension.
In contrast, the product metrics are language-dependent because their definition and implementation change across languages. 
For this reason, the \texttt{LanguageMetricsExtractor} sub-classes must implement that method.

Finally, the method \texttt{to\_csv()} saves the dataset on the disk, albeit the user can programmatically access it.
Besides, \texttt{LanguageMetricsExtractor}s must implement the \texttt{ignore\_file()} method to discard files written in a language different than the one considered for the analysis.

\section{RepoMiner: Usage Template}\label{sec:template}

\noindent \textsc{RepoMiner} can be installed through the Python Package Index as follows:

\begin{verbatim}
    pip install repository-miner
\end{verbatim}


\noindent or, alternatively, from the source code root folder:


\begin{verbatim}
    git clone https://github.com/radon-h2020/radon-repository-miner
    cd radon-repository-miner
    pip install .
\end{verbatim}

The user can import and use the \texttt{repominer} module in Python as shown below, where \texttt{language} and \texttt{Language} have to be replaced by one of the supported languages.
A concrete usage example is shown in \Cref{sec:case_study}.

  





\begin{figure}
    \centering
    \includegraphics[width=\linewidth]{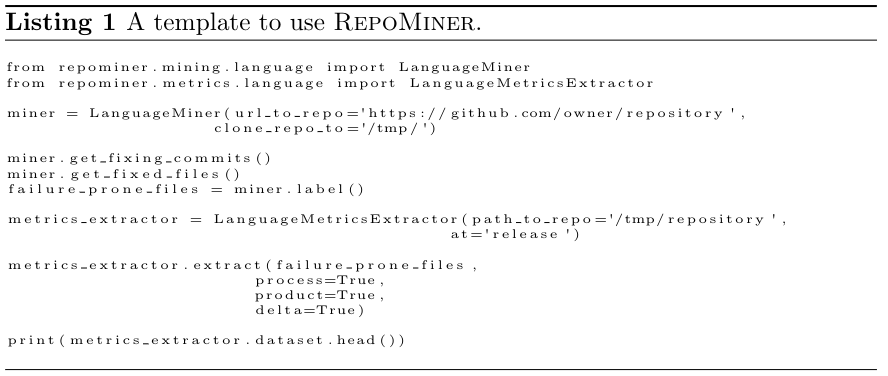}
    \caption{A template to use \textsc{RepoMiner}.}
\end{figure}

\section{The RADON Case Study for IaC Defect Prediction}\label{sec:case_study}

This section showcases how \textsc{RepoMiner} can be practically used and extended to different languages.
\textsc{RepoMiner} was developed under the European Union’s Horizon research and innovation project called RADON.\footnote{\url{https://radon-h2020.eu/}} 
The project aims at pursuing a broader adoption of serverless computing technologies within the European software industry. 
One of RADON’s key pillars is quality assurance tools for Infrastructure-as-Code (IaC), the DevOps principle for the automated configuration management of systems infrastructure~\cite{morris2016}.
IaC allows for solving issues related to manual configuration management using automatic provision and configuration of infrastructural resources based on practices from software development through the definition of machine-readable files.

On the one hand, these practices ensure consistent and repeatable routines for system provisioning and configuration changes.
On the other hand, frequent changes to the infrastructure code can inadvertently introduce defects~\cite{rahman2018characterizing} whose impact might be significant. 
For example, the execution of a defective IaC file erased the directories of about 270 Wikimedia users in 2017.\footnote{\url{https://wikitech.wikimedia.org/wiki/Incident_documentation/20170118-Labs}}
Our research contributes to this pillar by the development of a defect predictor to support the correctness of the infrastructure code developed in Ansible and Tosca, the most used IaC language in industry~\cite{guerriero2019adoption} and the language-agnostic standard for IaC\footnote{\url{https://www.oasis-open.org/committees/tc_home.php?wg_abbrev=tosca}}, respectively.

Because of their novelty and the lack of tools, in the two years of the project, we developed a tool suite (i) to automatically gather meaningful data about infrastructure code from open-source software repositories; (ii) extract process and code metrics from the mined IaC scripts; (iii) support DevOps engineers in training and evaluating defect prediction models, thus identifying snapshots of files likely to contain defects.
In this context, \textsc{RepoMiner} acts as a proxy between (i-ii) and (iii). Given a software repository, it executes points (i-ii) and generates a dataset of observations, representing a file at a given point of the repository's history. 
Each observation is a set of metrics extracted in (ii), metadata, and information about its failure-proneness (\ie the file is \textit{failure-prone} or \textit{neutral}).
A defect predictor can then use the dataset as the ground truth to train prediction models of failure-prone infrastructure code; thus, addressing point (iii).

\begin{figure}[htp]
    \centering
    \includegraphics[width=0.5\linewidth]{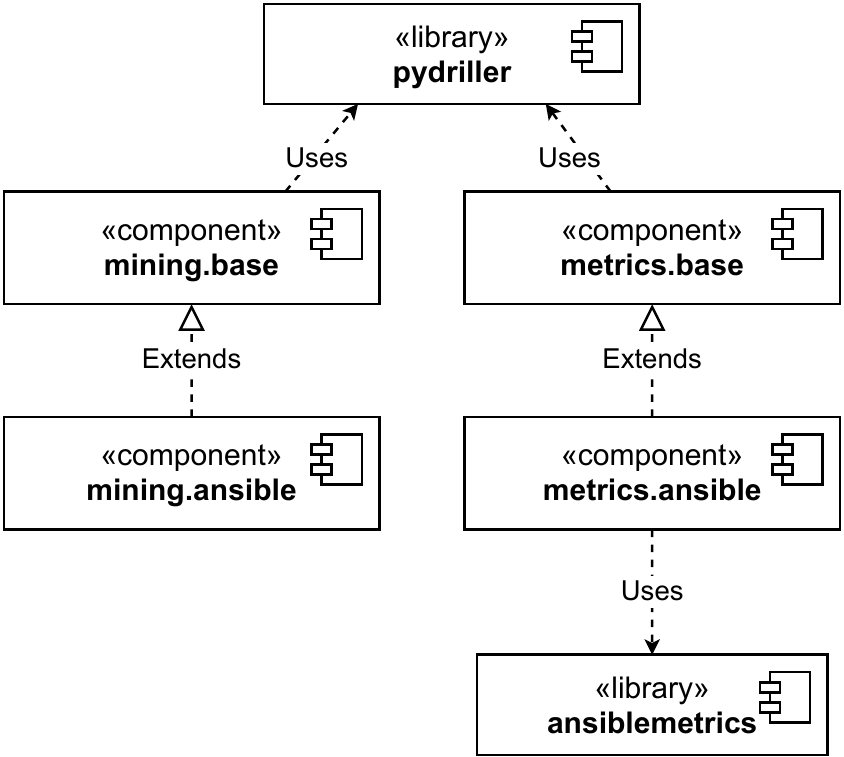}
    \caption{UML component diagram for \textsc{RepoMiner} in RADON.}
    \label{fig:concrete_component_diagram}
\end{figure}

\begin{figure}[htp]
    \centering
    \includegraphics[width=0.7\linewidth]{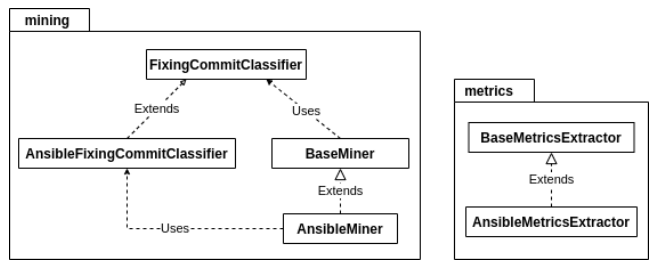}
    \caption{UML class diagram for \textsc{RepoMiner} in RADON.}
    \label{fig:concrete_class_diagram}
\end{figure}

For example, in our previous work on IaC defect prediction~\cite{DallaPalma2021tse}, we instantiated \textsc{RepoMiner} to collect failure-prone data to train defect prediction models for Ansible. 
The work aimed at helping software practitioners prioritize their inspection efforts for IaC scripts by proposing prediction models of failure-prone IaC scripts and investigating the role of product and process metrics for their prediction.
To that end, the tool was updated as in \Cref{fig:concrete_component_diagram}.
Two components were added specifically for Ansible, namely \texttt{mining.ansible} and \texttt{metrics.ansible}.
Therefore, the \texttt{LanguageMiner} and \texttt{LanguageMetricsExtractor} classes in \Cref{fig:class_diagram} were replaced by \texttt{AnsibleMiner} and \texttt{AnsibleMetricsExtractor} as in \Cref{fig:concrete_class_diagram}.

Extending the \texttt{BaseMiner} was effortless: only the methods \texttt{ignore\_file()} and \texttt{discard\_undesired\_commits()} had to be implemented.
For example, to analyse an Ansible repository, only files with \textit{.yml} extension  and under the directories \texttt{playbooks}, \texttt{meta}, \texttt{tasks}, \texttt{handlers}, and \texttt{roles} are analyzed.\footnote{Based on \url{https://docs.ansible.com/ansible/2.3/playbooks_best_practices.html#directory-layout} (Accessed on April 2021).}
In \texttt{discard\_undesired\_fixing\_commits()}, commits that do not modify Ansible files are removed from the list of defect-fixing commits.

Likewise, \texttt{AnsibleMetricsExtractor} extends the base class by implementing \texttt{ignore\_file()} and \texttt{get\_product\_metrics()}. 
However, it relies on the external library \textit{AnsibleMetrics}~\cite{DallaPalma2020softX} to calculate product metrics.\footnote{\url{https://github.com/radon-h2020/radon-ansible-metrics}}

Having made such modifications, we ran \textsc{RepoMiner} on 104 carefully selected Ansible projects in order to collect data for defect prediction. 
Below is a usage example of \textsc{RepoMiner} on one of the projects, based on the template shown in \Cref{sec:template}, and an extract of the output.  

    
  
  
\begin{figure}
    \centering
    \includegraphics[width=\linewidth]{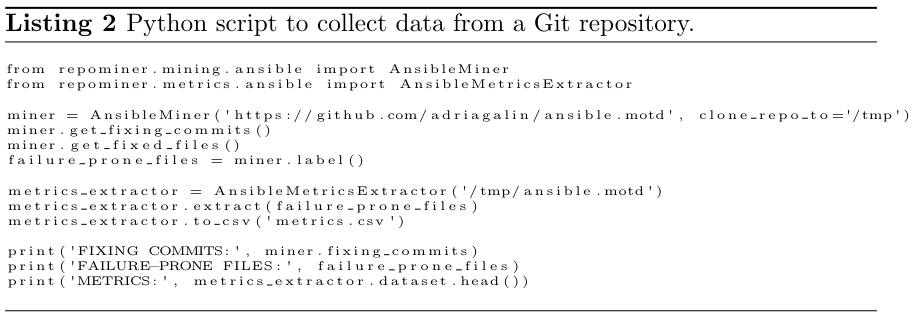}
    \caption{Python script to collect data from a Git repository.}
    \label{listing:concrete_usage}
\end{figure}

    


\begin{figure}
    \centering
    \includegraphics[width=\linewidth]{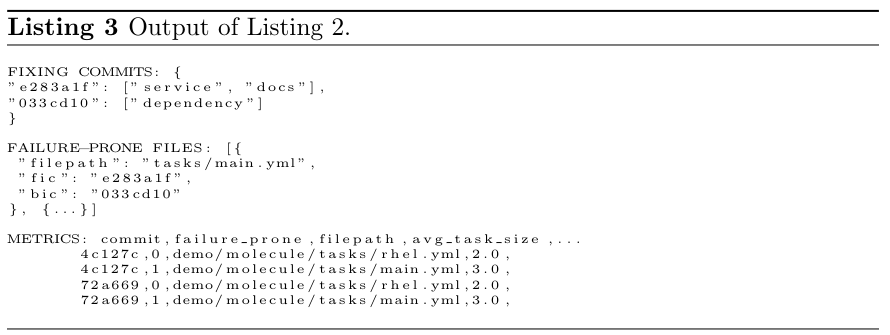}
    \caption{Output of \Cref{listing:concrete_usage}.}
\end{figure}

\begin{table}[htp]
    \centering
    \footnotesize
    \begin{tabular}{lrrrrr}
        & \textbf{Min} & \textbf{Mean} & \textbf{Std} & \textbf{Median} & \textbf{Max} \\\toprule
    \texttt{get\_fixing\_commits()} & 10 & 110 & 170 & 50 & 1170 \\\midrule
    \texttt{get\_fixed\_files()} & 0 & 30 & 40 & 10 & 300 \\ \midrule
    \texttt{label()} & 0 & 40 & 90 & 10 & 740 \\ \midrule
    \texttt{extract\_metrics()} & 20 & 450 & 1,400 & 170 & 13,540\\\midrule
    \textbf{Total} & 30 & 620 & 1,615 & 250 & 15,030\\
    \bottomrule
    \end{tabular}
    \caption{Statistics on execution time in seconds, rounded to the nearest ten.}
    \label{tab:performance}
\end{table}

\textsc{RepoMiner} identified almost five thousand fixing-commits (with a median of 26 per project) and approximately 14,000 defective observations (a median of 530 per project and three observations per release on average). 
Performance varied from a median of 50 seconds for identifying fixing commits to ten seconds for labeling failure-prone files and three minutes for extracting process and product metrics. Sometimes \texttt{extract\_metrics()} ran up to three hours before completing. 
It is the case of the repository \textit{redhat-openstack/infrared} where \textsc{RepoMiner} collected a dataset of approximately 76,000 observations over 6,000 commits divided in 145 releases. Therefore, product metrics had to be extracted for every observation and process metrics for every release.
\Cref{tab:performance} reports statistics related to the execution time of \textsc{RepoMiner} on the 104 projects. 
To measure the soundness of \textsc{RepoMiner} in identifying \textit{defect-fixing} commits of Ansible files, we uniformly selected and manually assessed a statistically relevant sub-sample of 357 fixing commits, and obtained a precision of 74\%.
We applied the same procedure to evaluate its precision in identifying \textit{bug-introducing} commits.
We validated 354 bug-introducing commits and obtained a precision of 84\%.


\section{Limitations and Extensions}\label{sec:limitation_extensions}

\textsc{RepoMiner} identifies fixing commits based on the defect taxonomy proposed by Rahman \etal~\cite{Rahman2020taxonomy}, which focuses on Infrastructure-as-Code.
However, other approaches could be implemented, combined, and compared.
For example, Babii \etal~\cite{babii2021mining} proposed \textsc{BOHR}, a repository where researchers can contribute heuristics to label Software Engineering data, such as bug-fixing commits.\footnote{\url{https://github.com/giganticode/bohr}} \textsc{BOHR} uses state-of-the-art weak supervision techniques to combine these heuristics and train classifiers operating on Software Engineering data.
Ideally, many similar heuristics might be suitable and easily integrated into \textsc{RepoMiner}, being both the tools written in Python.

\begin{figure}
    \centering
    \includegraphics[width=\linewidth]{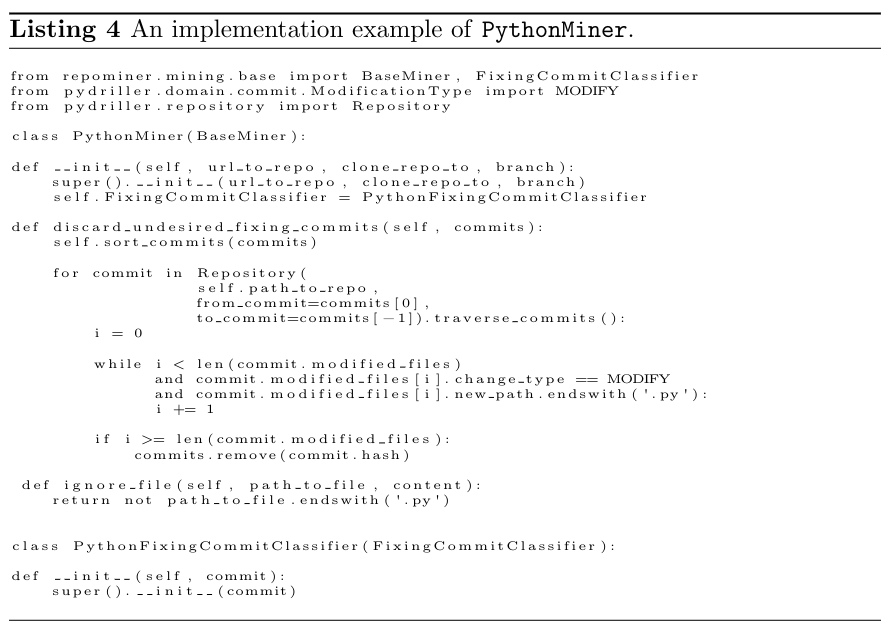}
    \caption{An implementation example of \texttt{PythonMiner}.}
    \label{fig:python_miner}
\end{figure}

Furthermore, although \textsc{RepoMiner} was initially developed to support Ansible and Tosca, it was designed to be easily extendable to additional Infrastructure-as-Code languages (\eg Chef and Puppet) and application code languages (\eg Python).
Let us suppose a developer would like to extend RepoMiner to support Python. They should implement \texttt{PythonMiner} and \texttt{PythonMetricsExtractor} to extend the respective base classes. 
The methods \texttt{discard\_undesired\_commits()} and \texttt{ignore\_file()} in \texttt{PythonMiner} would ensure that only commits fixing defects in Python files are kept and that the final dataset consists of solely Python files.
The code in \Cref{fig:python_miner} shows a possible implementation of these methods.





          
            




\begin{figure}
    \centering
    \includegraphics[width=\linewidth]{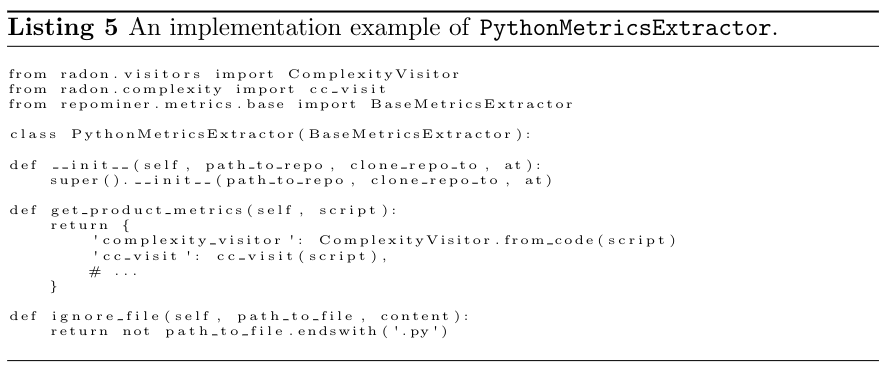}
    \caption{An implementation example of \texttt{PythonMetricsExtractor}.}
    \label{fig:python_metrics}
\end{figure}

The method \texttt{get\_product\_metrics()} in \texttt{PythonMetricsExtractor} would ideally rely on an external tool for obtaining raw metrics from Python code, such as \textit{radon}\footnote{Note, this tool is not related to the HORIZON 2020 RADON project. Docs available at \url{https://radon.readthedocs.io/en/latest/api.html} (Accessed on April 2021)}, as shown in \Cref{fig:python_metrics}.






\section{Conclusions}\label{sec:conclusions}
This paper presented \textsc{RepoMiner}, a novel tool to collect \textit{defect-fixing} commits, \textit{failure-prone} files, and their process and source code metrics.
It aims to make a step forward to support research when mining software repositories to ease the creation of datasets for defect prediction studies. 
\textsc{RepoMiner} is designed to work on \textit{git} repositories and potentially could be used for any infrastructure and application code language, such as Python and Java.

We illustrated the architecture of the tool and provided usage examples of its programming interface.
Nevertheless, further interaction modalities with the tool, for example, a web application, are planned for future work. 
In addition, we received encouraging feedback from the industrial partners within the EU's project \textsc{RADON}.
Therefore, we hope and plan to enlarge the user community around \textsc{RepoMiner} through their collaboration and dissemination.
We are confident that the proposed tool will help researchers and practitioners collecting meaningful data to feed defect prediction models either in the context of infrastructure and application code.

\section*{Acknowledgements}
This work is supported by the European Commission grant no. 825040 (H2020 RADON).

\balance

\bibliographystyle{elsarticle-num}
\bibliography{main}

\begin{thebibliography}{10}
\expandafter\ifx\csname url\endcsname\relax
  \def\url#1{\texttt{#1}}\fi
\expandafter\ifx\csname urlprefix\endcsname\relax\def\urlprefix{URL }\fi
\expandafter\ifx\csname href\endcsname\relax
  \def\href#1#2{#2} \def\path#1{#1}\fi

\bibitem{chaturvedi2013tools}
K.~K. Chaturvedi, V.~Sing, P.~Singh, {Tools in Mining Software Repositories},
  in: 2013 13th International Conference on Computational Science and Its
  Applications, IEEE, 2013, pp. 89--98.

\bibitem{Spadini2018}
D.~Spadini, M.~Aniche, A.~Bacchelli, {PyDriller: Python Framework for Mining
  Software Repositories}, in: Proceedings of the 2018 26th ACM Joint Meeting on
  European Software Engineering Conference and Symposium on the Foundations of
  Software Engineering, ESEC/FSE 2018, Association for Computing Machinery, New
  York, NY, USA, 2018, p. 908–911.

\bibitem{MSRtools2013}
K.~K. {Chaturvedi}, V.~B. {Sing}, P.~{Singh}, {Tools in Mining Software
  Repositories}, in: 2013 13th International Conference on Computational
  Science and Its Applications, 2013, pp. 89--98.

\bibitem{Kuhn2009}
A.~{Kuhn}, {Automatic Labeling of Software Components and Their Evolution Using
  Log-likelihood Ratio of Word Frequencies in Source Code}, in: 2009 6th IEEE
  International Working Conference on Mining Software Repositories, 2009, pp.
  175--178.

\bibitem{Gousios2009}
G.~Gousios, D.~Spinellis, {A Platform for Software Engineering Research}, in:
  Proceedings of the 2009 6th IEEE International Working Conference on Mining
  Software Repositories, MSR '09, IEEE Computer Society, USA, 2009, p. 31–40.

\bibitem{Rahman2020taxonomy}
A.~{Rahman}, E.~{Farhana}, C.~{Parnin}, L.~{Williams}, {Gang of Eight: A Defect
  Taxonomy for Infrastructure as Code Scripts}, in: 2020 IEEE/ACM 42nd
  International Conference on Software Engineering (ICSE), 2020, pp. 752--764.

\bibitem{kim2006automatic}
S.~{Kim}, T.~{Zimmermann}, K.~{Pan}, E.~J. {Jr. Whitehead}, {Automatic
  Identification of Bug-Introducing Changes}, in: 21st IEEE/ACM International
  Conference on Automated Software Engineering (ASE'06), 2006, pp. 81--90.

\bibitem{morris2016}
K.~Morris, {Infrastructure as Code}, O'Reilly Media, 2016.

\bibitem{rahman2018characterizing}
A.~Rahman, L.~Williams, {Characterizing defective configuration scripts used
  for continuous deployment}, in: 2018 IEEE 11th International Conference on
  Software Testing, Verification and Validation (ICST), IEEE, 2018, pp. 34--45.

\bibitem{guerriero2019adoption}
M.~Guerriero, M.~Garriga, D.~A. Tamburri, F.~Palomba, Adoption, support, and
  challenges of infrastructure-as-code: Insights from industry, in: 2019 IEEE
  International Conference on Software Maintenance and Evolution (ICSME), IEEE,
  2019, pp. 580--589.

\bibitem{DallaPalma2021tse}
S.~{Dalla Palma}, D.~{Di Nucci}, F.~{Palomba}, D.~A. {Tamburri},
  {Within-Project Defect Prediction of Infrastructure-as-Code Using Product and
  Process Metrics}, IEEE Transactions on Software Engineering.

\bibitem{DallaPalma2020softX}
S.~{Dalla Palma}, D.~{Di Nucci}, D.~A. Tamburri, {AnsibleMetrics: A Python
  library for measuring Infrastructure-as-Code blueprints in Ansible},
  SoftwareX 12.

\bibitem{babii2021mining}
H.~Babii, J.~A. Prenner, L.~Stricker, A.~Karmakar, A.~Janes, R.~Robbes, Mining
  software repositories with a collaborative heuristic repository, in: 2021
  IEEE/ACM 43rd International Conference on Software Engineering: New Ideas and
  Emerging Results (ICSE-NIER), IEEE, 2021, pp. 106--110.

\end{thebibliography}

\pagebreak
\appendix

\end{document}